\documentclass[aps,prl,twocolumn,showpacs,floatfix,groupedaddress,amsmath,amssymb]{revtex4}
\usepackage{graphicx}
\usepackage{mathrsfs}
\usepackage{array}


\newcommand{\ket} [1] {| #1 \rangle}

\makeatletter

\newcommand{\eref}[1]{Eq.~\ref{#1}}
\newcommand{\erefs}[2]{Eqs.~\ref{#1}-\ref{#2}}
\newcommand{\fref}[1]{Fig.~\ref{#1}}

\newcommand{\nsym}{\mbox{\tiny sym}}

\newcommand{\AKLT}{\mbox{\tiny AKLT}}
\newcommand{\nprod}{\mbox{\tiny prod}}

\usepackage{array}

\begin{document}
\title
{Symmetry protected entanglement renormalization}
\author{Sukhwinder Singh}
\affiliation{Center for Engineered Quantum Systems; Dept. of Physics \& Astronomy, \\Macquarie University, 2109 NSW, Australia}
\author{Guifre Vidal}
\affiliation{Perimeter Institute for Theoretical Physics, Waterloo, Ontario, N2L 2Y5, Canada}

\begin{abstract}
Entanglement renormalization is a real-space renormalization group (RG) transformation for quantum many-body systems. It generates the multi-scale entanglement renormalization ansatz (MERA), a tensor network capable of efficiently describing a large class of many-body ground states, including those of systems at a quantum critical point or with topological order. The MERA has also been proposed to be a discrete realization of the holographic principle of string theory.
In this paper we propose the use of symmetric tensors as a mechanism to build a symmetry protected RG flow, and discuss two important applications of this construction. First, we argue that symmetry protected entanglement renormalization produces the proper structure of RG fixed-points, namely a fixed-point for each symmetry protected phase. Second, in the context of holography, we show that by using symmetric tensors, a global symmetry at the boundary becomes a local symmetry in the bulk, thus explicitly realizing in the MERA a characteristic feature of the AdS/CFT correspondence.
\end{abstract}


\maketitle


Renormalization \cite{RG} is fundamental to our understanding of quantum many-body systems.
The renormalization group (RG) explores how the behavior of an extended system depends on the scale of observation. Important concepts such as universality, criticality or stability of phases are then explained in terms of the existence of fixed-points of the RG flow. On the other hand, in holographic string theory constructions \cite{Holo,AdSCFT}, an additional dimension corresponding to scale (or RG flow) is used to regard a many-body system as the boundary dual of a higher dimensional, bulk theory.

\begin{figure}
  \includegraphics[width=8.0cm]{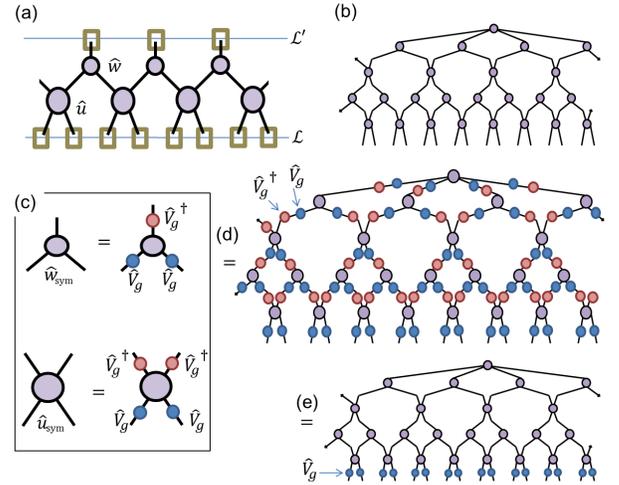}
\caption{
(a) Coarse-graining transformation, made of disentanglers $\hat{u}$ and isometries $\hat{w}$, that maps a lattice $\mathcal{L}$ into a coarse-grained lattice $\mathcal{L}'$.
(b) The MERA is obtained by repeatedly applying the coarse-graining transformation.
(c) Constraints satisfied by $\mathcal{G}$-symmetric disentanglers and isometries, \eref{eq:uwsym}.
A global symmetry $\mathcal{G}$ can be enforced by using $\mathcal{G}$-symmetric tensors: equality (b)=(d) follows from the symmetry constraints in (c); (d)=(e) is obtained using $\hat{V}_g^{\dagger}\hat{V}_g = \hat{\mathbb{I}}$ on each bond index, which only leaves $\hat{V}_g$ acting on the open indices; therefore (b)=(e), which amounts to \eref{eq:PsiSym}.
\label{fig:MERA}}
\end{figure}

Both renormalization and holography can be realized non-perturbatively, on the lattice, using \textit{entanglement renormalization}, a real-space coarse-graining transformation for many-body wave-functions \cite{ER,WFR}. A key aspect of entanglement renormalization is the removal of short-range entanglement using so-called disentanglers \fref{fig:MERA}(a). This leads to an efficient description of many-body states in terms of the multi-scale entanglement renormalization ansatz (MERA) \cite{MERA}, a tensor network that spans an additional dimension corresponding to scale, \fref{fig:MERA}(b). The MERA has been applied to the exploration of frustrated antiferromagnets \cite{MERAanti}, interacting fermions \cite{MERAfermion}, topological order \cite{MERAtopo, MERAanyon}, quantum criticality \cite{MERAcrit,MERACFT} and, more recently, holography \cite{Swingle,MERAHolo}.

In this paper we propose the use of symmetric tensors as a way to protect a global symmetry during the RG flow, and present two applications of this construction in the context of renormalization and holography. Specifically, we will explain how to use symmetric tensors to: (i) obtain the proper structure of RG fixed-points corresponding to symmetry-protected phases; and (ii) obtain a holographic description where a global symmetry at the boundary turns into a local symmetry in the bulk.

\textit{Symmetry protected entanglement renormalization.---}
Let $\mathcal{L}$ be a lattice made of $L$ sites, each described by a vector space $\mathbb{V}$, and let $\hat{V}_g:\mathbb{V}\rightarrow \mathbb{V}$ be a (unitary or anti-unitary) \textit{linear} representation of the group $\mathcal{G}$,
\begin{equation}\label{eq:group}
    \hat{V}_f \hat{V}_{g} = \hat{V}_{f\cdot g},~~~~~~~~~\forall f,g \in \mathcal{G}.
\end{equation}
We say that a many-body wave-function $\ket{\Psi} \in \mathbb{V}^{\otimes L}$ on the lattice $\mathcal{L}$ is invariant under the global symmetry $\mathcal{G}$, or $\mathcal{G}$\textit{-symmetric} \cite{covariant}, if
\begin{equation}\label{eq:PsiSym}
    (\hat{U}_g)^{\otimes L}\ket{\Psi} = \ket{\Psi}.
\end{equation}
Similarly, we say that a tensor is $\mathcal{G}$-symmetric if it has (unitary or anti-unitary) linear representations of $\mathcal{G}$ on all of its indices, and is invariant under its action. In the MERA \cite{MERA}, where the disentanglers $\hat{u}$ and isometries $\hat{w}$, see \fref{fig:MERA}(a), implement maps between one or two sites,
\begin{equation}\label{eq:uw}
    \hat{u}: \mathbb{V}\otimes \mathbb{V} \rightarrow \mathbb{V}\otimes \mathbb{V}, ~~~~~ \hat{w}: \mathbb{V} \rightarrow \mathbb{V} \otimes \mathbb{V},
\end{equation}
invariance under $\mathcal{G}$ is expressed as (see also \fref{fig:MERA}(c))
\begin{eqnarray}\label{eq:uwsym}
    \hat{u}_{\nsym} &=& \left(\hat{V}_g \otimes \hat{V}_g \right)\hat{u}_{\nsym}  \left(\hat{V}_g^{\dagger} \otimes \hat{V}_g^{\dagger} \right),\\
    \hat{w}_{\nsym} &=& \left(\hat{V}_g \otimes \hat{V}_g \right)\hat{w}_{\nsym}  \left(\hat{V}_g^{\dagger} \right).
\end{eqnarray}
The result of coarse-graining a $\mathcal{G}$-symmetric wave-function $\ket{\Psi}$ using $\mathcal{G}$-symmetric disentanglers $\hat{u}_{\nsym}$ and isometries $\hat{w}_{\nsym}$ is also a $\mathcal{G}$-symmetric state, and we thus obtain a symmetry protected RG flow. In particular, \fref{fig:MERA}(b,d,e) shows that the MERA, which is built by concatenating several coarse-graining transformations, represents then a $\mathcal{G}$-symmetric state $\ket{\Psi}\in \mathbb{V}^{\otimes L}$ \cite{Singh}.

\begin{figure}[t]
  \includegraphics[width=8.5cm]{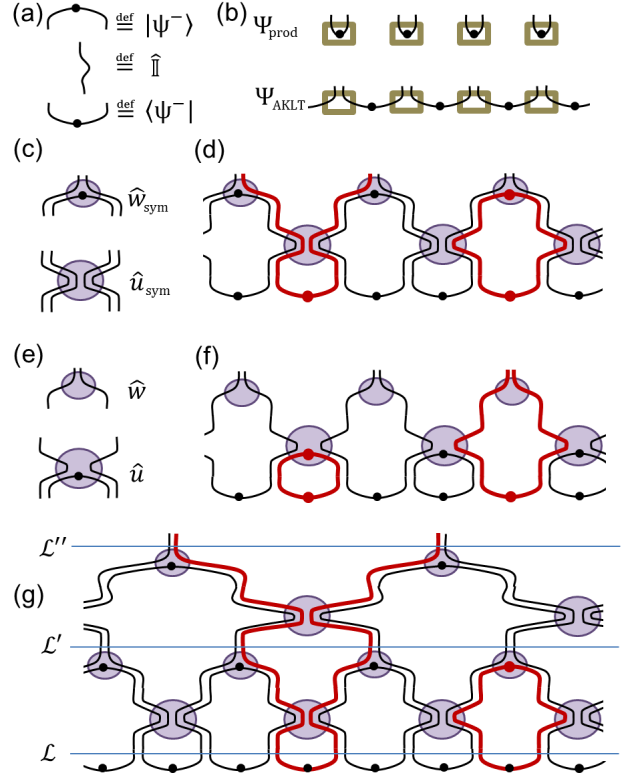}
\caption{
(a) Pictorical representation of the singlet state $\psi^{-}$ (in bra and ket forms) and of the identity operator $\hat{\mathbb{I}}$ on the space $\mathbb{C}_2$ of a spin-$\frac{1}{2}$ degree of freedom.
(b) States $\Psi_{\nprod}$ and $\Psi_{\AKLT}$, \erefs{eq:prod}{eq:AKLT}, made of singlet states $\psi^{-}$.
(c) $\mathbb{Z}_2^T$-symmetric disentanglers and isometries. Notice that $\hat{u}_{\nsym}$ is the identity operator (it cannot remove short-range entanglement, which is protected by the symmetry).
(d) As a result, the coarse-graining transformation only succeeds at removing every second singlet state $\ket{\psi^-}$. In red are examples of a singlet state that cannot be removed (left) and one that can be removed (right) during coarse-graining.
(e) Non-symmetric disentanglers and isometries. By allowing some of their indices to contain projective representations of $\mathbb{Z}_2^T$, instead of linear representations, disentanglers can now eliminate short-range entanglement. Notice that in this case each lower index of $\hat{w}$ (each upper index of $\hat{u}$) only contains one line, corresponding to a vector space $\mathbb{V}' \cong \mathbb{C}_2$ instead of $\mathbb{V}\cong \mathbb{C}_4$ in Eq. \ref{eq:uw}.
(f) As a result, the state $\Psi_{\AKLT}$ becomes the state $\Psi_{\nprod}$ after just one coarse-graining transformation.
(g) The ground state $\Psi_{\AKLT}$ is thus a fixed-point of the RG flow \textit{with} symmetry protection.
\label{fig:Z2fixed}}
\end{figure}

\textit{Application I: symmetry protected RG flow.---} For concreteness, let us specialize to time reversal symmetry, $\mathcal{G} = \mathbb{Z}_2^{T}$, acting on a lattice $\mathcal{L}$ where each site is described by two spin-1/2 degrees of freedom, denoted $\vartriangleleft$ (left) and $\vartriangleright$ (right) spins, $\mathbb{V} \cong (\mathbb{C}_2)_{\vartriangleleft} \otimes (\mathbb{C}_2)_{\vartriangleright}$. On each spin, time reversal acts anti-unitarily by means of operator $\hat{\mathcal{T}}$,
\begin{equation}\label{eq:T}
    \hat{\mathcal{T}} \equiv i \hat{\sigma}_y \hat{\mathcal{K}}, ~~~~~~~\hat{\sigma}_{y}\equiv \left(\begin{array}{cc} 0 & -i \\ i & 0 \end{array} \right),
\end{equation}
where $\hat{\sigma}_y$ is a Pauli matrix and $\hat{\mathcal{K}}$ denotes complex conjugation. Since $\hat{\mathcal{T}}\hat{\mathcal{T}}=-\hat{\mathbb{I}}$, time reversal acts on a single spin-$\frac{1}{2}$ as a \textit{projective} representation of $\mathbb{Z}_2$. However, when acting simultaneously on two spins, we have $(\hat{\mathcal{T}}^{\otimes 2})(\hat{\mathcal{T}}^{\otimes 2})=(-\hat{\mathbb{I}})^{\otimes 2} = (\hat{\mathbb{I}})^{\otimes 2}$, and therefore time reversal acts on each site of lattice $\mathcal{L}$ by means of a \textit{linear} representation of $\mathbb{Z}_2$. A \textit{singlet state} $\ket{\psi^{-}} \equiv \frac{1}{\sqrt{2}}(\ket{01} - \ket{10})$ of two spins is invariant under time reversal, as follows from $(i\hat{\sigma}^y)^{\otimes 2} \ket{\psi^{-}} = \ket{\psi^{-}}$. We then consider the following two $\mathbb{Z}^{T}_2$-symmetric many-body states, \fref{fig:Z2fixed}(a,b),
\begin{eqnarray}\label{eq:prod}
    \ket{\Psi_{\nprod}} &\equiv& \bigotimes_{s\in \mathcal{L}} \ket{\psi^{-}}_{s,\vartriangleleft;s,\vartriangleright}, ~~~\\
\label{eq:AKLT}
    \ket{\Psi_{\AKLT}} &\equiv& \bigotimes_{s\in \mathcal{L}} \ket{\psi^{-}}_{s,\vartriangleright;s+1,\vartriangleleft}.
\end{eqnarray}
In the \textit{product} state $\ket{\Psi_{\nprod}}$ each site $s$ of $\mathcal{L}$ is in a singlet state $\ket{\psi^{-}}$, and therefore there is no entanglement between different sites. Instead, in the \textit{entangled} state $\ket{\Psi_{\AKLT}}$, which is related to the AKLT state \cite{AKLT}, a singlet state entangles the right spin of site $s$ with the left spin of site $s+1$, and therefore there is entanglement between nearest neighbor sites. It is easy to see that both wave-functions are unique ground states of $\mathbb{Z}^{T}_2$-symmetric, local, gapped Hamiltonians, namely $\hat{H}_{\nprod} \equiv \sum_{s\in \mathcal{L}} \hat{\vec{\sigma}}_{s,\vartriangleleft} \cdot \hat{\vec{\sigma}}_{s,\vartriangleright}$ and $\hat{H}_{\AKLT} \equiv \sum_{s\in \mathcal{L}} \hat{\vec{\sigma}}_{s,\vartriangleright} \cdot \hat{\vec{\sigma}}_{s+1,\vartriangleleft}$. [Indeed, each interaction $\hat{\vec{\sigma}} \cdot \hat{\vec{\sigma}}$ acts on a different pair of spins, and $\ket{\psi^{-}}$ is the unique ground state of $\hat{\vec{\sigma}} \cdot \hat{\vec{\sigma}}$].

States $\ket{\Psi_{\nprod}}$ and $\ket{\Psi_{\AKLT}}$ are interesting because in a system invariant under time inversion they belong to two different phases \cite{Pollmann, Chen}, with $\ket{\Psi_{\AKLT}}$ corresponding to the Haldane phase \cite{AKLT}. Following Refs. \cite{Pollmann, Chen, Schuch}, we say that two local, gapped Hamiltonians correspond to the same phase of matter if we can smoothly deform one into the other by a means of a path of local, gapped Hamiltonians. Similarly, we say that two ground states are in the same phase if they are the unique ground states of those Hamiltonians. In absence of symmetry, in one spatial dimension there is only one phase \cite{Pollmann, Chen, Schuch}, and we can choose the product ground state $\ket{\Psi_{\nprod}}$ as its representative. However, when we add the restriction that a global, on-site symmetry $\mathcal{G}$ must be preserved at all times, then one obtains several inequivalent phases, called symmetry protected topological (SPT) phases, classified by the second group cohomology  $H^{2}(\mathcal{G},$U$(1))$ of the symmetry group $\mathcal{G}$ \cite{Pollmann, Chen, Schuch}. For time reversal symmetry there are only two SPT phases, because the second group cohomology of $\mathbb{Z}_2^T$  is $\mathbb{Z}_2$, $H^{2}(\mathbb{Z}^{T}_{2}, $U$(1)) = \mathbb{Z}_2$, and $\ket{\Psi_{\nprod}}$ and $\ket{\Psi_{\AKLT}}$ are their representatives.

Our goal is to use symmetry protected entanglement renormalization to reproduce the classification of SPT phases from the perspective of the RG flow. We have found that, indeed, states $\ket{\Psi_{\nprod}}$ and $\ket{\Psi_{\AKLT}}$ correspond to two inequivalent fixed points of the RG flow when $\mathbb{Z}_2^T$-symmetric disentanglers and isometries are used in the coarse-graining transformation; and that they both flow to the same fixed-point if non-symmetric tensors are used. The RG flow of $\ket{\Psi_{\nprod}}$ is trivial: a product state is transformed into a product state \cite{product}, and we will not discuss it here. Let us focus instead on state $\ket{\Psi_{AKLT}}$. \fref{fig:Z2fixed}(c),(d) shows a specific choice of $\mathbb{Z}_2^T$-symmetric disentangler and isometry $(\hat{u}_{\nsym}, \hat{w}_{\nsym})$ that coarse-grain state $\ket{\Psi_{AKLT}}$ into a state locally identical to $\ket{\Psi_{AKLT}}$. In other words, $\ket{\Psi_{AKLT}}$ is a fixed-point of the symmetry protected RG flow, \fref{fig:Z2fixed}(g). In contrast, when non-symmetric disentanglers and isometries are allowed, we can find a pair $(\hat{u}, \hat{w})$ that coarse-grains $\ket{\Psi_{AKLT}}$ into the product state $\ket{\Psi_{\nprod}}$ after just one step of coarse-graining, \fref{fig:Z2fixed}(e),(f).

It is important to emphasize that although the tensors $(\hat{u}, \hat{w})$ in \fref{fig:Z2fixed}(e) are not $\mathbb{Z}_2^T$-symmetric (some of their indices carry projective representations of $\mathbb{Z}_2^T$), the resulting coarse-graining transformation in \fref{fig:Z2fixed}(f) is fully $\mathbb{Z}_2^T$-symmetric. In other words, it is only when we attempt to implement the $\mathbb{Z}_2^T$-symmetric coarse-graining transformation in \fref{fig:Z2fixed}(f) by means of \textit{local} transformation (disentanglers and isometries) that we are forced to use projective representations and thus non-symmetric tensors.

A byproduct of the above analysis is an exact MERA representation of $\ket{\Psi_{\AKLT}}$ in terms of the pair of fixed-point, $\mathbb{Z}_2^T$-symmetric tensors $(\hat{u}_{\nsym}, \hat{w}_{\nsym})$. This construction can be extended to an arbitrary group $\mathcal{G}$, by replacing the singlet state $\ket{\psi^-}$ of two spin-$\frac{1}{2}$ degrees of freedom with an entangled state of two (generalized) spins that is invariant under the action of the suitable projective representations of $\mathcal{G}$ acting on the two spins, and thus obtain RG fixed points and exact MERA representations for each element of $H^{2}(\mathcal{G},$U$(1))$. Previously, exact fixed-point MERA representations were only known for topologically ordered phases \cite{MERAtopo}.

The above analysis is also reassuring from a numerical perspective. Symmetric tensors are commonly used in MERA algorithms, because they provide several computational advantages, including faster implementations algorithms and access to quantum numbers \cite{Singh}. To these advantages now we add the ability to identify in which SPT phase a given symmetric Hamiltonian $\hat{H}$ is. For instance, one can numerically optimize a symmetry protected, scale-invariant MERA (including a few transitional coarse-graining transformations until the RG fixed-point is reached) with the algorithm of Ref. \cite{MERACFT}, and determine the SPT phase by analyzing the resulting fixed-point pair $(\hat{u}_{\nsym}, \hat{w}_{\nsym})$ \cite{scaleinvariantMERA}.

\begin{figure}[t]
  \includegraphics[width=8.5cm]{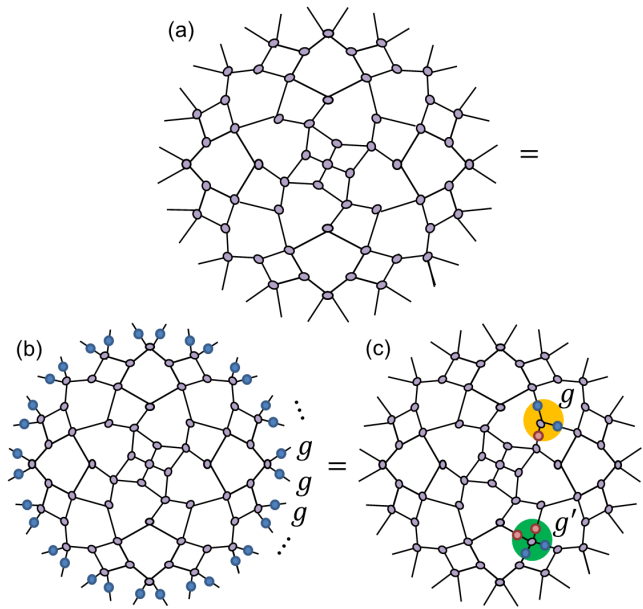}
\caption{
(a) MERA for the state $\ket{\Psi}$ of a lattice $\mathcal{L}$ with $L=32$ sites in $d=1$ space dimensions. This holographic tensor network expands one additional dimension, with the radial direction corresponding to scale.
The MERA can be used to represent the ground state of a quantum critical system, corresponding to a CFT, and its network of tensors is a discrete version of AdS geometry.
(b) Use of $\mathcal{G}$-invariant disentanglers and isometries, \eref{eq:uwsym}, guarantees that the boundary state $\ket{\Psi}$ has $\mathcal{G}$ as a global symmetry, \eref{eq:PsiSym}.
(c) It also implies that $\mathcal{G}$ acts as a local symmetry in the bulk of the tensor network
(with its local action given precisely by \eref{eq:uwsym}). Indeed, the MERA is invariant under the transformation of e.g. two tensors by two different group elements $g$ and $g'$.
\label{fig:Holo}}
\end{figure}

\textit{Application II: symmetry protected holography.---} The AdS/CFT correspondence \cite{AdSCFT} asserts the equivalence between a conformal field theory (CFT) in $d+1$ space-time dimensions and a theory of gravity in anti-de Sitter (AdS) space-time in $d+2$ dimensions, where the additional dimension in AdS corresponds to changes of scale in the CFT. This correspondence establishes a dictionary between properties of the CFT (seen as living at the boundary of AdS) and properties of the gravity theory (seen to live in the bulk of AdS). In particular, the scaling dimensions of operators in the CFT are related to masses of fields in the gravity theory, and a global symmetry in the boundary CFT translates into a local symmetry in the bulk \cite{AdSCFT}.

Being based on a RG transformation, the MERA represents the ground state of a local lattice Hamiltonian in $d$ spatial dimensions by a tensor network spanning $d+1$ dimensions, where the additional dimension also corresponds to changes in scale. [Notice that the description of a time-independent ground state only requires $d$ space dimensions ($d+1$ in holography), instead of the $d+1$ space-time dimensions ($d+2$ in holography) required to study time-dependent properties]. For instance, \fref{fig:Holo}(a) shows a MERA for the ground state of a spin chain, that is, a lattice model in $d=1$ space dimensions, and it spans $d+1=2$ dimensions. This tensor network can efficiently describe the ground state of a quantum critical system, which in the continuum may correspond to a CFT. As a matter of fact, in that case one can extract the conformal data of the CFT from the MERA \cite{MERACFT}. Suggestively, the scaling dimensions of operators are obtained from the masses that control the exponential decay of correlations in the scale direction (as extracted from the eigenvalues of a \textit{scaling super-operator} that implements changes of scale), which matches one of the ingredients of the holographic dictionary. In addition, the scaling of entanglement entropy in MERA was originally computed by adding contributions from different scales \cite{MERA0}, in close analogy with the holographic computation of entanglement entropy in a CFT \cite{HoloEntropy}. Ref. \cite{Swingle} proposed that, indeed, the MERA should be interpreted as a lattice realization of the holographic principle, with the tensor network reproducing a discrete version of the AdS geometry. Since then, this proposal has been supported and extended by several authors \cite{MERAHolo}. Here, we aim to provide further support for this interpretation by proposing how to explicitly realize in the MERA another entry of the holographic dictionary, namely the one that relates global symmetries at the boundary with local symmetries in the bulk.

We have argued above, see \fref{fig:MERA}, that a global symmetry $\mathcal{G}$ of the state $\ket{\Psi} \in \mathbb{V}^{\otimes L}$ can be incorporated into the MERA by using $\mathcal{G}$-symmetric tensors. In the holographic interpretation of the MERA, the global symmetry is a property of the boundary theory, see \fref{fig:MERA}(a),(b). However, using $\mathcal{G}$-symmetric tensors also implies that the tensor network is invariant under the action of $\mathcal{G}$ on individual tensors, where it acts simultaneously on all the indices of a tensor, see eq. \ref{eq:uwsym}. In other words, the bulk theory contained in the tensor network of the MERA has acquired a local symmetry $\mathcal{G}$, meaning that it is invariant under the action of $\mathcal{G}$ on different tensors by different group elements, such as $g$ and $g'$ in \fref{fig:Holo}(c).

In most practical applications of the AdS/CFT correspondence, one exploits the fact that it is a weak-strong duality: when, say, the bulk theory is weakly coupled, and therefore can be treated within perturbation theory, then the boundary theory is strongly coupled (and vice-versa). It is important to realize that the MERA, being based only on the existence of an RG flow, offers an efficient holographic representation of most known ground states \cite{branching}, regardless of whether either the bulk or boundary theories are weakly coupled, or even whether the boundary theory is related to a CFT. For any such ground state with a global symmetry $\mathcal{G}$, we can now obtain a holographic description with an explicit local realization of the symmetry $\mathcal{G}$ in the bulk.

\textit{Discussion.---} Previously, symmetric tensors had been used in the MERA as a means to guarantee the exact implementation of a global symmetry in the resulting many-body state. This also had other advantages, including the ability to target specific quantum numbers during a simulation, and a significant reduction in computational space and time attained by exploiting the inner structure of $\mathcal{G}$-symmetric tensors \cite{Singh}. However, in principle there is no need to use symmetric tensors in order to enforce a global symmetry and, as a matter of fact, in some cases using non-symmetric tensors results in a simpler description \cite{SinghPrep}.
Here we have proposed the use of symmetric tensors as a means to build a symmetry protected renormalization group flow with the correct structure of fixed-points \cite{OtherRGSchemes}. We have also seen that this construction produces a holographic description where the global symmetry at the boundary is explicitly realized as a local symmetry in the bulk. Although we have illustrated these ideas in the context of one dimensional systems, our results apply also to higher dimensions, where one may encounter a richer variety of phases. In particular, we expect that symmetry protected entanglement renormalization will provide an RG framework to study the interplay between symmetry protection and topological order \cite{Mesaros13,Essin13}.

Entanglement renormalization, as a non-perturbative, real-space, lattice realization of the powerful ideas of the renormalization group, offers a unique testing-ground to explore various aspects of renormalization and the holographic principle. In this paper, by investigating how to implement a global symmetry in entanglement renormalization, we hope to have contributed to our understanding of the role played by symmetries in the renormalization group flow and in holographic descriptions of many-body states.

\textbf{Acknowledgements.-} The authors thank R.C. Myers, W. Witczak-Krempa, and N. Jones for useful discussions. G.V. thanks the Australian Research Council Centre of Excellence for Engineered Quantum Systems. This research was supported in part by Perimeter Institute for Theoretical Physics. Research at Perimeter Institute is supported by the Government of Canada through Industry Canada and by the Province of Ontario through the Ministry of Research and Innovation.

Note: At the final stages of preparation of this manuscript we became aware of the recent paper by C.-Y. Huang, X. Chen, and F.-L. Lin on ``Symmetry Protected Quantum State Renormalization" \cite{SPWFR}. \textit{Quantum state renormalization} \cite{WFR} and \textit{entanglement renormalization} \cite{ER} appear to be equivalent concepts. Accordingly, their symmetry-protected versions, as addressed in Ref. \cite{SPWFR} and here respectively, are also equivalent.


\begin{thebibliography}{74}
\bibitem{covariant} The present discussion can be generalized to states that transform under $(\hat{V}_g)^{\otimes L}$ as an arbitrary linear representation of the group $\mathcal{G}$.
\bibitem{product} Strictly speaking, for each site it would be enough to keep a one-dimensional subspace $\tilde{\mathbb{V}} \subseteq (\mathbb{C}_2)_{l} \otimes (\mathbb{C}_2)_{r}$ corresponding to the singlet state $\ket{\psi^-}$.
\bibitem{scaleinvariantMERA}
If the symmetry group $\mathcal{G}$ only has a small number of SPT phases, one can try all possible fixed-point pairs $(\hat{u}_{\nsym}, \hat{w}_{\nsym})$ corresponding to these phases, for which we have described an exact expression, and restrict the numerical optimization (energy minimization) to the transitional layers.
\bibitem{branching} In some cases, such as systems in $d\geq 2$ spatial dimensions with a $d-1$ dimensional Fermi surface, an efficient description may require using a generalization of the MERA known as branching MERA \cite{branching2}.
\bibitem{OtherRGSchemes} Symmetry protected phases had been previously discussed in terms of a symmetry protected RG scheme in Ref [19]. We emphasize that the RG scheme employed in Ref [19] is fundamentally different from that employed in our paper. In particular, the RG scheme of Ref [19] has no disentanglers and, as a result, states that only differ in short range details flow to different fixed points.
\end{thebibliography}
\end{document}